\documentstyle[aps,twocolumn]{revtex}

\def\be{\begin{eqnarray}}
\def\ee{\end{eqnarray}}
\def\a{\alpha}
\def\b{\beta}
\def\d{\delta}
\def\D{\Delta}

\def\g{\gamma}

\def\f{\varphi}
\def\l{\lambda}
\def\o{\omega}

\def\ddg{d^{\dagger}}

\def\dg{\dagger}
\def\s{\sigma}
\def\t{\theta}
\def\ur{\uparrow}
\def\dr{\downarrow}
\def\e{\epsilon}
\def\z{\zeta}

\begin{document}
\title{ Time reversal symmetry breaking in cuprates \\induced
by the spiral spin order.}
\author{ M.Ya.Ovchinnikova}
\address{Joint Institute of Chemical Physics of RAS,
Kosygin str.,4, 117334, Moscow, Russia}
\wideabs{
\maketitle
\begin{abstract}
We propose a new interpretation of the spontaneous time
reversal symmetry breaking (TRSB) observed recently in a pseudogap state of
cuprates (Kaminsky et al.). It is shown that the TRSB dichroism in ARPES
signal may be connected with the local spin spiral structures in system.
It may be caused by a spin-orbit interaction and by spin polarization of
electrons at various sections of Fermi surface in spiral state. Angular
dependence of dichroism signal is studied in schematic KKR approximation.
Tests are proposed to check an existence of the local spiral spin structure
and to distinguish it from the TRSB state with micro-currents constructed by
Varma.
\end{abstract}

Pacs: 71.10.Fd,  74.20.Rp, 74.20.-z

}

%
%
 The nature of a pseudogap (PS) state of high-$T_c$ cuprates in underdoped
(UD) region remains the intriguing problem \cite{1,2}.
Recently, using the angular-resolved photoemission (ARPES) with circularly
polarized light (CPL), Kaminsky et al. \cite{3} reveal new property of
the pseudogap state of UD $Bi_2Sr_2CaCu_2O_{8-\d}$ (BSCCO). It was shown that
this state displays a spontaneous time-reversal symmetry breaking (TRSB).
Earlier Varma \cite{4} had predicted the possibility of TRSB in
cuprates. They proposed the fascinate ground state  with circular
microcurrents inside the plaquettes of $CuO_2$ plane with definite alignment
of orbital angular momenta associated with these micro-currents.
Namely, the up-directed orbital momenta arrange along one
diagonal and the down-directed orbital momenta arrange
along other diagonal. The proposed in \cite{4,5} alignment of orbital
angular momenta is not connected with any spin alignment.

The aim of present paper is to discuss an alternative possibility for
constructing the state with TRSB. Here we propose a state in which the TRSB
is due to a spiral spin structure. The arguments in favor of such
hypothesis  are following. The electric field of CPL really interacts only
with the orbital motion. Therefore the TRSB dichroism implies a definite
orientation of orbital angular momenta $<L_n>\neq 0$. Such momenta, centered
on the atoms, may be induced by aligned spin momenta $<S_n>\neq 0$ through
the spin-orbit interaction. This means that in ARPES one could observe a TRSB
dichroism $D$ if the photoemission setup can selectively measure the ejected
electrons with definite spin projection $\s=\ur$ or $\dr$. Then a sign of the
TRSB dichroism would depend on a sign of $\s$. Since usually the ARPES
is non-selective with respect to the final spin projection of an
ejected electron, the summary TRSB dichroism is expected to be zero if a
mean spin polarization of initial states is zero. But for the
spiral spin structure the occupancy $n_{k\s}$ of initial one-electron band
state $\{k\s\}$ with definite $k$ depends on $\s$. Such spin polarization of
initial $k$-state may induce the non-zero TRSB effects in ARPES signal. So we
calculate a dichroism which might manifest in ARPES in case of the spiral
spin configuration of cuprates.

The scheme of experiment \cite{3} is given in Fig.~1. The right (left)
polarized light with a propagation vector ${\bf q}_{\g}$ impacts the
crystal surface determined by a normal vector ${\bf n}$. The
$xz$ plane is one of the mirror planes of crystal with x-axis along
the $CuO$ bonds or along a diagonal direction. The
ejected electron has a final momentum ${\bf k}_f$. The ARPES
intensity $I\sim |M_{if}|^2 \d(E_i-E_f-\hbar \o)$ is determined by a matrix
element of interaction $O=(e/2m_e c){\bf (A p+pA)}$ with field
\be
M=A_{\a}F_{\a}; ~~~F_{\a}=<\psi_f |p_{\a}|\psi_i(k)>
\label{1}
\ee
between initial and final states. In dipole approximation it contains the
vector potential ${\bf A}$ of the right ($\z_R=1$) or left ($\z_L=-1$)
CPL with complex amplitude
\be
{\bf A_{R(L)}}=A_0 [{\bf e}_x \cos{\t_\g}+i\z_{R(L)}{\bf e}_y+
{\bf e}_z \sin{\t_\g}]
\label{2}
\ee
For given configuration of the setup vectors ${\bf n},~{\bf k}_f,~
{\bf q}_\g$ the ARPES dichroism signal $D$ is determined by a relative
difference of intensities for both light polarizations:
\be
D=(M_R-M_L)/(M_R+M_L)
\label{3}
\ee
We study the symmetry properties of ARPES matrix elements with respect to
reflection in a mirror plane of a crystal, which is perpendicular to the
surface in a typical photoemission experiment. Following \cite{3},
consider first a time reversal invariant initial state $\psi_i (k)$ and let
${\bf q}_\g$ and ${\bf n}$ lie in the mirror plane $m$ of crystal (here the
$xz$ plane). Then, the dichroism signal $D$ is nonzero only if ${\bf k}_f$
does not lie in the mirror plane $m$ and D has an opposite signs for $k$ at
different sides of mirror plane. Such dichroism is called a geometrical
one. This large effect has been observed at any doping \cite{3}. But in UD
BSCCO the residual dichroism $(D\neq 0)$ has been observed even for coplanar
configuration of ${\bf n}$, ${\bf k}_f$, ${\bf q}_\g$, in which all three
vectors lie in mirror plane $m$. Further we adopt a definitions
$q_z$,${\bf q}$ and $k_z$, ${\bf k}$  for the normal and 2D intra-layer
components of the photon and final electron momenta ${\bf q}_{\g}$  and
${\bf k}_f$ correspondingly.

Consider first a large  geometrical dichroism and then discuss a possible
origin of the observed residual dichroism connected with TRSB of the ground
state of UD cuprate. We suggest that the main contributions to matrix element
give the space regions inside the atomic spheres. This is in accordance with
a fact that the frequency dependencies  of photoemission intensity roughly
repeat the dependencies of the photoemission cross-sections coming from
corresponding atomic components \cite{6}.

The formalism for evaluating the optical matrix element for general lattice
within KKR scheme is given in \cite{7}. Some corrections should be introduced
to provide the common asymptotic $\sim e^{ik_f r}$ of the final wave function
of the ejected electron outside the sample ($z>0$). We restrict our
consideration by the one-step model (see \cite{8}) describing a coherent
part of photoemission. Incorporation of rescattering and relaxation
processes in frame of generalization to a three-step model are needed to
describe the background in the energy distribution function of ejected
electron. We believe that one-step model is sufficient to describe in
qualitative manner the angular dependence of dichroism. For this aim we will
use the most simplified form of the initial and final states in the
process.

The starting point in calculating of $F_\a$ in (1) is the KKR wave function
for a multicomponent lattice \cite{7}. In Hartree-Fock representation the
one-particle initial state $\psi_i$ with quasi-momentum ${\bf k}$
is a superposition of orbitals belonging to each center
\be
\psi_i(r,k)={1\over \sqrt{N}} \sum_{n,\b} B(n_z)
e^{ikr}i^lC_{L\b}^i\psi_{L\b}^i(r-R_{n\b})
\label{4}
\ee
Here $\b$ enumerates all atoms placed at $R_{n\b}$ in unit cell
$n=(n_x,n_y,n_z)$ and $L={l,m}$ are the angular momentum quantum numbers of
orbitals inside each atomic sphere $|r-R_{n\b}|<a_{n\b}$. For the upper
valence anti-bonding band of $CuO_2$ plane the main orbitals $\psi_{L\b}^i$
are the $d_{x^2-y^2}$ orbital of Cu and $p_x,~p_y$ orbitals of two oxygens
$O_x,O_y$.  These orbitals constitute a basis set of the Emery model.

Thus the initial one-electron state is taken as superposition of the main
real orbitals and in the second quantization representation  it is
\be
\psi_i(k,\s)=\sum_{n_z} B_i (n_z) [c_d d^{\dg}_{k\s} +ic_x x^{\dg}_{k\s}
+ic_y y^{\dg}_{k\s}]
\label{5}
\ee
The corresponding site operators  $d^{\dg}_{n\s}$, $x^{\dg}_{n,\s}$,
$y^{\dg}_{n,\s}$ of Emery model refer to the real functions
$d_{x^2-y^2}(r-R_{n,d})$, $p_{\nu}(r-R_{n,\nu })$ with
$R_{n,\nu}=R_n+{\bf e}_{\nu}a/2$, $\nu=x,y$.
The functions are considered to extend
inside the corresponding atomic (muffin-tin) spheres.  The real coefficients
$c_d$, $c_x$, $c_y$ in (5) are obtained from solution of the Emery model.
In Eq.(4,5) the term under the sum refer to a layer number $n_z$.
The amplitudes $B(n_z)$ depending on distance of layer from the surface
describe in phenomenological manner a coherent or incoherent interlayer
transport along $z$ near the surface depending on the phase correlations
between different layers. For standard bulk initial state $\psi_i$ used in
\cite{7} $B(n_z)\sim \exp{(ik_z n_z)}$.

A final state inside the sample is taken in similar KKR form with the same
in-plane momentum $k$:
\be
\psi_f={1\over \sqrt{N}} \sum_{n,\b} B^f (n_z)
e^{ikR_{n\b}}i^l C_{L\b}Y_{L}^*\psi_{L\b}(r-R_{n\b})
\label{6}
\ee
Here each of functions $\psi_{L\b}$ with angular momentum quantum numbers
$L=(l,m)$ is determined inside atomic spheres around corresponding
center $R_{n,\b}$. Influence of surface at z=0 is described by introducing
the factors $B^f (n_z)$, by  phases $\d_{l,\b}$  of  complex
coefficients
\be
C_{L\b}=|C_{L\b}| e^{\d_{l\b}},
\label{7}
\ee
and by explicit angular spherical harmonics $Y_{L}=Y_{lm}({\hat k}_f)$
depending on direction of final momentum $k_f$. The phases are specific for
centers $\b$ in unit cell and for angular momentum $l$. These phases arise
from a matching of final state (6) inside the sample with the
common plane wave $\sim e^{ik_fr}$ in empty space outside it.
The phase modulation of contributions in (6) determines the geometrical
dichroism of the photoemission.

The origin of spherical harmonics in Eq.(6) and of the phase modulation of
coefficients (7) may be illustrated by next consideration.  First let us
construct the final state $\psi_{f,\b} (r)$ for the electron photoemission
into direction ${\hat k}_f$ from one center $\{n,\b\}$ only.  According to
\cite{9} it must be such function of continuum which has an
asymptotic form of plane wave $\sim e^{i{\bf k}_f {\bf r}}$ and incoming
radial waves at $|r-R_{n\b}|\to \infty$.  This final
state is
\be
\psi_f=\exp{(ik_f R_{n\b})}\sum_{lm} i^l e^{i\d_{l\b}}
Y_{lm}({\hat r})Y_{lm}^*({\hat k}_f)\f_{l}(r_{\b})
\label{8}
\ee
where $r_\b=|r-R_{n\b}|$. The scattering phase $\d_{l\b}$ for the orbital
momentum $l$ is defined by the asymptotic of real radial function of
continuum $\f_{l\b}(r)\sim {1\over r}\sin{(kr-\pi l/2+\d_{l\b})}$. (According
to the KKR approach one may consider that asymptotic form is achieved at the
surface of the muffin-tin sphere.)

In similar manner the final state $\psi_f$ for the electron ejected from
$N_s$ centers of the surface layer should be a function which at large $z>0$
have an asymptotic form with common plane wave $\sim e^{i{\bf k}_f {\bf r}}$
and incoming spherical waves contributed from different centers.

If $k_f|R_{n\b}-R_{n',\b'}|> 1$ and if we neglect the secondary scattering
processes, then inside each of nonoverlapping muffin-tin spheres surrounding
center $(n\b)$ of surface layer the final state wave function should just
have a form (6) with complex coefficients $C_{L'\b}\sim e^{i\d_{l\b}}$.
Really the secondary processes synchronize the phases $\d_{l\b}$ of all
contributions into the final state from different angular harmonics and
different centers.  The KKR bulk solution for the final state $\psi_f(k_f)$
found in \cite{7} takes into account the phase and amplitude synchronization
of all secondary processes, but it neglect the necessary additional
synchronization and phase modulation coming from the boundary surface where
solutions should be matched with the plane wave with momentum $k_f$ .

In order to study in qualitative manner the angular dependence of dichroism
and its symmetry, it is sufficient to use the form (6) of the final state
without specifying the values and phases of $C_{L'\b}$ in (6). So we use
Eqs.(5,6) for schematic representation of initial and final states to
study the symmetry and possible angular dependence of dichroism manifested in
ARPES. The components $F_{\a}$ of the matrix element (1) is expressed
as a sum of integrals inside atomic spheres around centers $(\b)$ for
corresponding channels $l\to l'$
\be
M_{R(L)}=A_{\a}(\z)F^{\nu}_{\a}(l'{\hat k})
\label{9}
\ee
Here $A_{\a}$ are components of the vector potential (2)  depending on the
right or left polarization of CPL, $\z=\z_{R(L)}=\pm 1$, and ${\hat k}={\bf
k}_f/k_f$. Functions $F^{\nu}_{\a}(l',{\hat k})$ correspond to
real initial orbitals $\nu =d_{x^2-y^2},p_x,p_y $. To obtain them we use the
selection rules $l'=l\pm 1$ for orbital angular momenta for integrals inside
atomic spheres.  For simplicity we retain only the matrix elements for the
following transitions $p_{x(y)}\to s,d$ and $d_{x^2-y^2}\to p$ from the
$O_{x(y)}$ and $Cu$ centers of $CuO_2$  plane. According the KKR calculations
\cite{7} such transitions give main contributions. Omitting of
transition $d_{x^2 -y^2}\to f$ at $Cu$ center does not change the symmetry
properties of the calculated dichroism. This leads only to neglecting
the small higher harmonics in angular dependence of the ARPES
intensity.

The resulting expressions for functions $F^{\nu}_{\a}(l'{\hat k})$ are
presented in Table.  For the $p\to s,d$ transitions in oxygen they
include also the factors
\be
g_{x(y)}=s_{x(y)}/\sqrt{s_x^2+s_y^2};~~ s_{x(y)}=\pm\sin{(k_{x(y)}/2)};
\label{10}
\ee
They originate from angular dependence of real amplitudes $c_{x(y)}$ and
$c_d$ of different orbitals in the initial band state (5) of Emery model
(with effective parameters $\e_d, \e_p,t_{pd},t_{pp}$).
At $t_{pp}\ll t_{pd}$ the amplitudes in (5) are
\be
c_{x(y)}=g_{x(y)}\sin{\eta};~~c_d=\cos{\eta};
\label{11}
\ee
$$
\tan{2\eta}={{2t_{pd}(\cos{k_x}+\cos{k_y})}/({\e_d-\e_p}}).
$$
Extension to large $t_{pp}$ does not change the symmetry of
amplitudes.

The coefficients $C_{0(I,II)}(k)$ in table include: 1) a sum over the
layers $\sum B^*_f(n_z)B_i(n_z)$ based on the phenomenological or
tight-binding dependencies $B(n_z)$; 2) the phase factors $\exp{(i\d_{l'})}$
coming from boundary conditions; 3) the reduced integrals $<l'\b\parallel
p\parallel l\b>$ over angular variables after removing the $m$ dependence; 4)
the radial integrals; 5) the factors $\sin{\eta},~\cos{\eta}$ from amplitudes
(11).

Then the ARPES dichroism signal is
\be
D(\f)=Im\{M(\D M )^*\}/(|M|^2+|\D M|^2)
\label{12}
\ee
where $M=M_R+M_L$, $\D M=M_R-M_L$ and angles $\t, \f$ describe a final
momentum ${\bf k}_f$ .  Dependence (12) may be presented as $D(\f)\sim
{\tilde G}(k)\sin{\f}$ with even function ${\tilde G}(k)$
relative to reflection in mirror plane zx. According to (9) the
quantities $M_{R(L)}$ are determined by complex constants $C_0, C_I,
C_{II}$ whereas the rest angular functions listed in Table are real
functions.  It can be shown that the dichroism signal is zero if all
coefficients $C_{L\b}$ in (7) have the same phases $\d_{\b,l}$ or their
differences are multiple to $\pi$.  So a representation of final state with
correct phases is significant for description of geometrical dichroism.

At ${\hat q}_\g={\bf n}$ when photon impacts normally to the $CuO_2$
plane one obtains
\be
\begin{array}{ll}
M=&C_{II}\sin{\t}\cos{\f}+g_x[C_0-C_I(\cos^2{\t}\\
&-\sin^2{\t}\cos{2\f})] +g_yC_I\sin^2{\t}\sin{2\f}\\
\end{array}
\label{13}
\ee
\be
\begin{array}{ll}
\D M=&-C_{II}\sin{\t}\sin{\f}+g_y[C_0-C_I(\cos^2{\t}\\
&+\sin^2{\t}\cos{2\f})] +g_xC_I\sin^2{\t}\sin{2\f}\\
\end{array}
\label{14}
\ee
Here $\t,\f$ are the polar angles of ${\bf k}_f$.
It is seen from (11,12) that the dichroism signal $D(\f)$ is odd
function $D(-\f)=-D(\f)$ and it go to zero at $\f=\pi$ or
$\f=0$. This is an expected property of geometrical dichroism.

Manifestations of geometrical dichroism depend on numerous parameters.
Fig.2 illustrates an examples of functions $D(\f)$ for three angles
$\t_g= 0,~ \pi/6,~\pi/3$ of the photon momentum in mirror plane $zx$ and
for $k$ moving along boundary $|k_x\pm k_y|=\pi$ and $k_z=|k|$.
We ascribe the following arbitrary values  for relative amplitudes
$|C_0/C_I|=|C_0/C_{II}|=1.0$ and the relative phases
$\{\d_{l'=0}(O),~\d_{l'=2}(O),~\d_{l'=1}(Cu)\}=\{0,~3\pi/4, \pi/4\}$
of coefficients $C_0,~C_I,~C_{II}$ in  different
channels of oxygen and Cu centers. Two setup configurations with $x$ along
$CuO$ bond or along diagonal direction are considered.
Function $D(\f)$ is odd function of $\f$
and it vanishes at $\f=0,\pi$.  The calculated geometrical dichroism
disappears for all $\f$ if all phase differences $\d_l-\d_{l'}=\pi m$ are
multiple to $\pi$.  This is just the case for the Cu- and O- contributions to
the matrix element calculated in \cite{7}. There the standard KKR bulk wave
functions were used and an additional phase modulation were neglected. At a
normal photon impact ($\t_{\g}=0$) a dichroism signal is zero on each mirror
plane of tetragonal lattice, i.e. at $\f={\pi \over 4}m$.

Now we take into account the spin-orbit interaction on Cu with a constant
$\l$
\be
V_{LS}=\l\sum_n {\bf L}_n {\bf S}_n
\label{15}
\ee
Then the initial band function $\psi^i_{k\s}$ transforms to
$\psi^i_{k\s}+\d\psi$ in a way equivalent to replacement of
$\ddg_{x^2-y^2,\s}$ in (5) by
\be
\ddg_{x^2-y^2,\s}+C_{\l}[2i\xi_{\s}\ddg_{xy,\s}-\xi_{\s}\ddg_{zx,-\s}
-i\ddg_{zy,-\s}]
\label{16}
\ee
in Eq.(5,4). Here $\xi_\s=\s/|\s|=\pm 1$ and  $C_l\sim
\l/2\d E$ where $\d E$ is the energy difference of the d-orbitals  of $x^2-y^2$
and $xy, yz,xz $ symmetries. Additional contribution to $\psi^i_{k\s}$ leads
to changes $M\to M+\d M$, $\D M\to \D M+\d\D M(\s)$ in Eqs.(10, 11,12).
The TRSB dichroism signal at the normal photon impact ($\t_{\g}=0$) is
determined then by a quantity
\be
\d\D M=\xi_{\s}4C_{\l}C_{II}\sin{\t}\cos{\f}
\label{17}
\ee
As a result the dichroism signal $D({\f,\s})$ of photoemission with the final
momentum $k_f$ and spin projection $\s$ of the ejected electron has a form
\be
D(\s,k)=A\sin{\f}-
{ {\xi_{\s}Re(MC_{II}^*)} \over {|M|^2+|\D M|^2} }
4C_{\l}\sin{\t}\cos{\f}
\label{18}
\ee
Here $M,\D M$ and $C_{II}$ are determined by Eqs.(13,14,7) and by functions
from Table.  Only linear in $\l$ term  is retained in (16). It is determined
only by admixture of $d_{xy}$ orbital in Eq.(18). The contributions from
d-orbitals of $xz,~ yz$ symmetries in (16) are of a 2-nd order of magnitude
in $\l$. The second term in (18) is even function of $\f$ and
have nonzero value at $\f=0$ or $\pi$ when all three vectors ${\bf q}_\g,
{\bf n}, {\bf k}_f$ lye in the mirror plane xz and geometrical dichroism
disappears.

Since a sign of $D(\f=0)$ depends on sign of spin projection $\s$ of
ejected electron, the overall dichroism $D=\sum_{\s}D(\s,\f)$ should be zero
for initial paramagnet (PM) state of system. So, for PM state the dichroism
at $\f=0$ or $\pi$ would be observed only if one selects the ejected
electrons with definite spin projection on ${\bf n}$.  For this PM state the
time reversal symmetry is broken just by a measurement of spin polarization
of photoelectron.

However, there are  the TRSB states in which the different regions of $k$
space are characterized by different spin polarization. For example, for the
ground state with a spiral spin structure the TRSB effect manifests
in ARPES by the nonzero overall dichroism at arrangement of all vectors
${\bf q}_\g, {\bf n}, {\bf k}_f$ in the mirror plane.

Let us demonstrate the polarization selectivity of the level occupancies in
$k$-space for the spiral state of the 2D $t-t'-U$ Hubbard model.
Calculations were carried out for model with $U/t=6,~~t'/t=0.1$ at doping
0.15 holes per site. The spiral mean-field (MF) solution is characterized by
average spins $<{\bf S}_n>= d({\bf e}_x \cos{Qn} + {\bf e}_y \sin{Qn})$
rotating in the $xy$ plane.
We study the MF states of two types - with the spirality vectors
\be
Q_I=(\pi-\d Q_x, \pi),~~
Q_{II}=(\pi -\d Q, \pi-\d Q)
\label{19}
\ee
along $CuO$ bond or along diagonal direction.
The spectral function $A_\s (k,\o )$ at $\o=0$ for definite spin
projection $\s$ on $z$- axis (perpendicular to the spin rotation plane) is
equal to \be A_\s (k,\g )=\sum_{if}|<\psi_f|c_{k\s}|\psi_i>|^2 f(E_i)
\d_{\g}(E_i-E_f )
\label{20}
\ee
Here the Fermi function $f(E_i)$ depends on the one-electron levels of MF
solution and $\d_{\g}(x)$ is the $\d$-function broadened with parameter
$\g\sim 0.05 t$.  Fig.3 presents an image of the spin-selective and overall
spectral functions $A_{\s=\ur}({\bf k})$ and $A({\bf k})=\sum_{\s}A_{\s}$ at
$\o=0$ for two types of spiral states.  Dark and light gray lines in Fig.3
correspond to main and shadow spin-selective ($\s=\ur$) sections of Fermi
surfaces. Similar images for $\s=\dr$ are obtained from those for $\s=\ur$ by
inversion ${\bf k}\to -{\bf k}$. The spin dependence of the level occupancy
in $k$ space is connected with the spin currents $J_{\ur}= J_{\dr}\sim Q$
existing in spiral state.

Thus, the TRSB state with spiral
spin structure certainly has a spin-selective sections of Fermi surface.
As a consequence, one can observe the TRSB dichroism of the ARPES signal even
at coplanar arrangement of the setup vectors ${\bf q}_\g, {\bf  n},
{\bf k}_f$ in mirror plane. Two factors are decisive here: 1) The ARPES
signal corresponds to a definite local region of $k$ which for given
spirality vector $Q$ is associated with definite spin polarization; 2)
According to Eq.(16,17) a definite spin polarization induces via the
spin-orbit interaction the orbital angular momenta on Cu centers and
corresponding nonzero dichroism at coplanar setup configuration.

In order to estimate effect we use the spin-orbit constant $\l\sim 800~
\verb"cm"^{-1}$ following from excitation spectrum of $Cu^I, Cu^{II}$
\cite{10} and
a splitting of d-orbitals $\d E=E_{x^2-y^2}-E_{xy}\sim 1\div 2$ eV in crystal
field. Then one has a value $C_{\l}=0.0025\div 0.005$ for amplitude in
(16). At setup configuration $\t_{\g}=0, ~\t=\t_k=\pi/4$, $\f=\pi$ or 0 and
at the same arbitrary chosen coefficients (7) as were used in Fig.2
($|C_I/C_0|=|C_{II}/C_0|=1,~~\d_{\b,l}=0,3\pi/4,\pi/4$), one obtains for the
TRSB dichroism  $max |D|=0.033\div 0.066$ at $k$ corresponding
$\f=0 \verb" or " \pi$  and lying in mirror plane. This value is consistent
the TRSB dichroism signal $\sim 3\div 5$\%  observed in UD cuprates \cite{3}.

In conclusion, it is shown here that the TRSB dichroism observed in the ARPES
spectra of the UD cuprates may be connected with a local spiral spin order in
system. This hypothesis differs from the model TRSB state proposed by
Varma et al. \cite{4,5} who connect the TRSB with aligned charge circular
microcurrents on plaquettes of $CuO_2$ plane. Instead, the spiral spin order
means appearance of the local spin currents ${\bf J}_{\ur}=-{\bf J}_{\dr}$
of the macro-scale, about the domain size. Existence of different domains
with different signs and values of the TRSB dichroism signal has been shown
by  study a set of samples of UD BSCCO in \cite{3}. The following test for
new hypothesis may be proposed. The rotation of sample on $180^{\circ}$
around z-axis does change a sign of the TRSB dichroism $D(\f=\pi)$ in our
hypothesis and does not change the sign in case of the TRSB state
constructed in \cite{4}. In the
former case the rotation changes the signs of the spin currents and of spin
polarization. Relative signs of the TRSB dichroism signal at different mirror
planes of cuprate for two types of spiral states (19) and for the state
proposed by Varma are illustrated in Fig.4.  These signs may be measured only
if the ARPES signal comes from the same domain of a sample before and after
its rotation and if the spin currents of spiral state are pined during
rotation of sample.  Note, that in case of the ferromagnet alighnment of
spins in the surface layers, the TRSB dichroism would have the same signs
at all directions in mirror planes of the lattice.

A great sensitivity of the Fermi surface (FS) to the spin structure puts the
questions important for understanding the pseudogap state of BSCCO:  Whether
the observed FS is a composed result coming from several domains with
different currents? What is a dynamics of these currents and domains in the
UD cuprates?  May the spin fluctuations be frosen near the surface into a
static domains with spiral or ferromagnet spin order?
Additional test for the supposed local spiral order is
possible.  One can measure the spin polarization $<{\bf S}>$ of electrons
ejected from different sections of the Fermi surface and to check the
correlations of direction of $<{\bf S}>$ with the sign of the TRSB dichroism
signal $D(\f)$ at $\f=\pi$. Such a program requires a spin-selective
detection of photoelectrons.

Work is supported by Russian Fund of Fundamental Research (Projects
No. 00-03-32981 and No. 00-15-97334). Author is grateful to
A.A.Ovchinnikov and V.Ya.Krivnov for useful discussions.


%

\begin{table}

\caption{
Functions $F_\a^\g(l',{\hat k})$ deter\-mining the $\a$-com\-ponents of matrix
element in Eq.(1). Index $\g$ numerates the contributions
originated from different orbitals of initial state: from $p_{x(y)}$
orbitals of oxygens $O_{x(y)}$ or $d_{x^2-y^2}, d_{xy}$ orbitals of $Cu$; final
channels $s,p,d$ correspond to angular momenta $l'=0,1,2$.  Functions
$G_{x(y)},~L_k,~g_x$  and $g_y,~P_k$ are even and odd functions relative to
mirror planes zx correspondingly.
}
\begin{tabular}{lllll}
$\psi_{i\b}$&$\psi_f$&$F_x$&$F_y$&$F_z$\\
\tableline
$p_x$&$s$&$C_0 g_x$&0&0\\
$p_x$&$d$&$C_I g_x G_x$&$C_I g_x P_k$&
$C_I  g_x L_k$\\
$p_y$&$s$&0&$C_0 g_y$&0\\
$p_y$&$d$&$C_I g_y P_k$&$C_I g_y G_y$&
$C_I g_y L_k$\\
$d_{x^2-y^2}$&$p$&$C_{II}\sin{\t}\cos{\f}$&
$-C_{II}\sin{\t}\sin{\f}$&0\\
$d_{xy}$&$p$&$C_{II}\sin{\t}\sin{\f}$&
$C_{II}\sin{\t}\cos{\f}$&0\\
\end{tabular}
\tablenotetext[1]{$G_{x(y)}=\pm\sin^2{\t}\cos{2\f}-\cos^2{\t}+1/3$}
\tablenotetext[2]{$g_{x(y)}$ are determined by Eqs.(10)}
\tablenotetext[3]{$L_k=\sin{2\t}\cos{\f}$;~~
$P_k=\sin^2{\t}\sin{2\f}$}
\end{table}

%


\begin{figure}
\caption{
The setup configuration of the ARPES experiment [3]. The
propagation vector ${\bf q}_\g$ of CPL lies in a mirror plane xz
and ${\bf k}_f$, ${\bf k}$ are the final momentum of ejected electron and its
component in $CuO_2$ plane (xy plane).}
\label{fig1}
\end{figure}

\begin{figure}
\caption{
Dependence of geometrical dichroism $D(\f)$ on the azimuthal angle $\f$ of
vector ${\bf k}_f$ for k moving along the nesting lines $|k_x\pm k_y|=\pi$.
Solid, dashed and thin curves correspond to angles $\t_\g=0,~\pi/6,~\pi/3$ of
the photon momentum $q$. Setup configurations with x along diagonal
direction or along the $CuO$ bonds refer to left or right graphics. Arbitrary
taken coefficients (7) are given in text.}

\label{fig2}
\end{figure}
\begin{figure}
\caption{
The images of the spin-selective (plots a,c) and overall (plots b,d) spectral
functions $A_{\ur}(k,\o)$, $A(k,\o)$ at $\o=0$ in Brillouin zone. Main and
shadow Fermi surface are shown for the spiral states with the spirality
vectors $Q=Q_I$ or  $Q=Q_{II}$ from (19) (plots a,b or c,d correspondingly) .
}
\label{fig3}
\end{figure}
\begin{figure}
\caption{
Relative signs of the TRSB dichroism signal for ${\bf k}$ lying on
different mirror planes of lattice for spiral states with
spirality vectors $Q_I$, $Q_{II}$ from (19), or for state proposed by
Varma (VS). }
\label{fig4}
\end{figure}


\begin{references}
%
\bibitem{1}
T.Timusk, B.Statt, Rep.Progr.Phys. {\bf 62}, 61 (1999)
%
\bibitem{2}
J.C.Campusano, M.R.Norman, M.Raderia, in Physics of Conventional
and Unconventional Superconductors. ed.K.H.Benneman, J.B.Ketterson, 2002;
cond-mat/0209476.
%
\bibitem{3}
A.Kaminsky, S.Rosenkranz, H.M.Fretwell, J.C.Campuzano, Z.Li, H.Raffy,
W.G.Cullen, H.You, C.G.Olson, C.M.Varma and H.H\"ochst, E-print Archive,
cond-mat/0203133.
%
\bibitem{4}
C.M.Varma, Phys.Rev.B{\bf 61}, R3804 (2000).
%
\bibitem{5}
M.E.Simon and C.M.Varma, E-print Archive, cond-mat/0201036.
%
\bibitem{6}
Y.Sakisaka, J.\-Electr.\-Spectr.\-Relat.\-Phen.{\bf 66}, 387 (1994)
%
\bibitem{7}
M.Lindroos, S.Sahrakorpi and A.Bansil, E-print Archive cond-mat/0109039,
Phys.Rev.B{\bf 65}, 054514 (2002).
%
\bibitem{8}
A.Damascelli, Z.X.Shen and Z.Hussain, E-print Archive,
cond-mat/0208504, to be publish in Rev.Mod.Phys.
%
\bibitem{9}
L.Landau and L.Lifshitz, Quantum Mechanics. M.Nauka, 1989.
%
\bibitem{10}
Atomic Energy Levels, ed. R.F.Bacher, S.Goudsmoth, McGraw-Hill Book Company,
N.Y.-London, 1932.
%
\end{references}
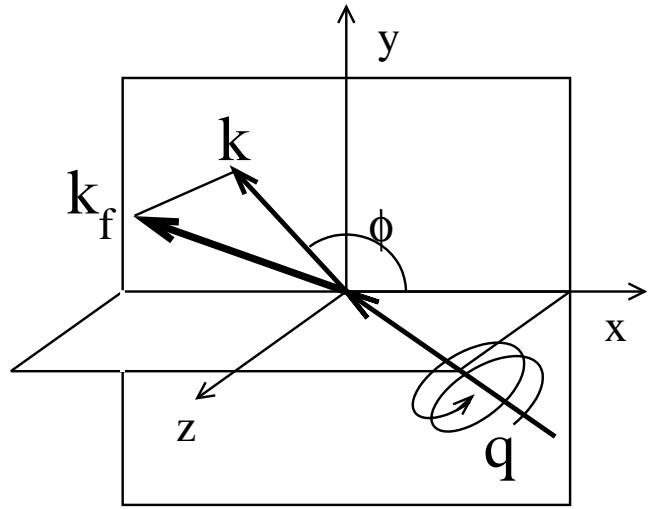
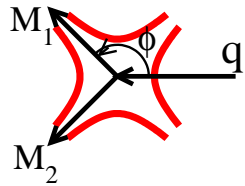
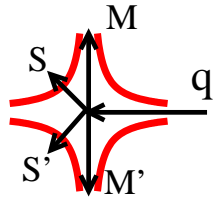
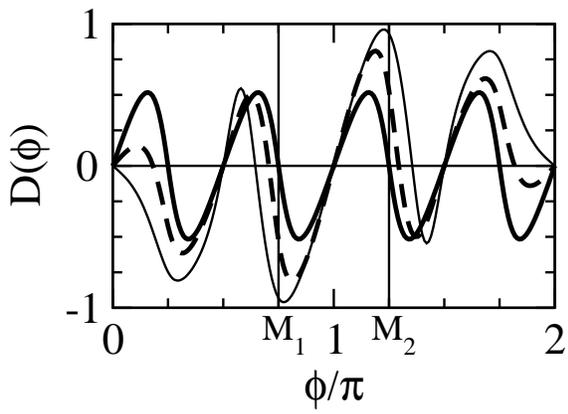
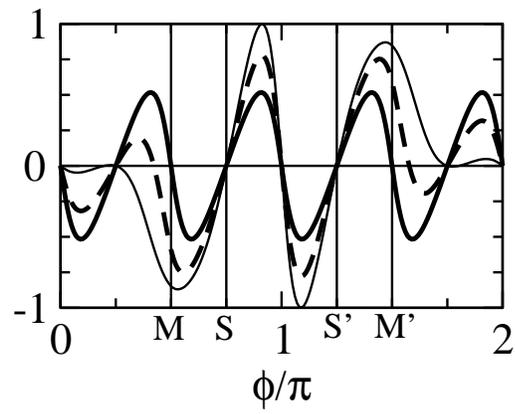
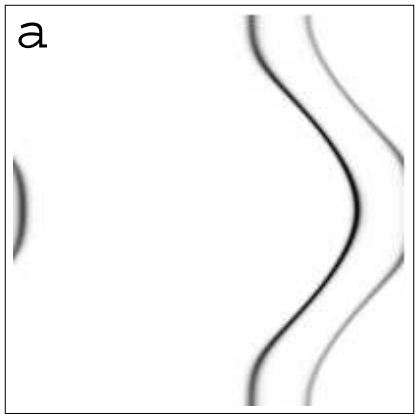
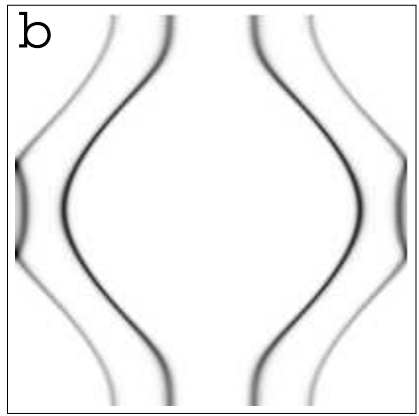
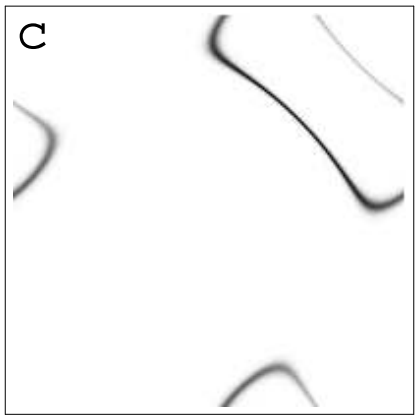
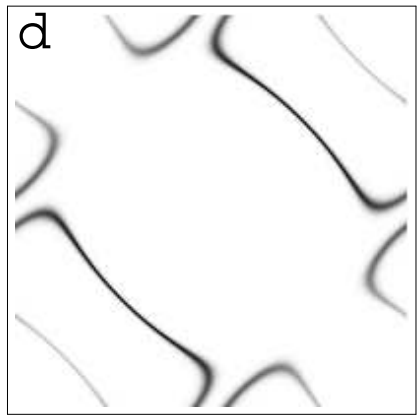
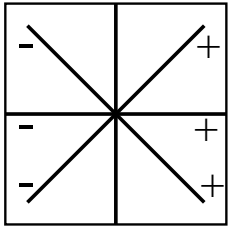
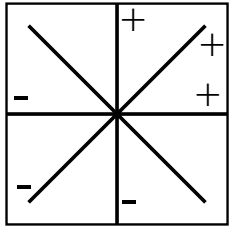
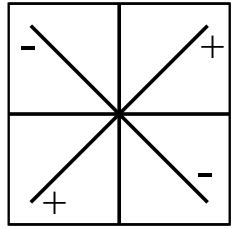
\end{document}